\newcommand{\rem}[1]{}
\documentclass[prd,twocolumn,showpacs,showkeys,showpacs,preprintnumbers,amsmath,amssymb]{revtex4-2}
\usepackage{amsfonts,amssymb,amsmath,mathrsfs}
\usepackage{url}
\usepackage{graphicx}
\newtheorem{thrm}{Theorem}%[section]
\newtheorem{lem}[thrm]{Lemma}
\newtheorem{prop}[thrm]{Proposition}

\begin{document}
%%%%%%%%%%%%%%%%%%%%%%%%%%%%%%%%%%%%%%%%%%%%%%%%%%%%%%%%%
\title[Cotton gravity]{Friedmann equations in the Codazzi parametrization\\ 
of Cotton and extended theories of gravity\\ 
and the Dark Sector} 
\author{Carlo Alberto Mantica}
\email{carlo.mantica@mi.infn.it} 
\author{Luca Guido Molinari} 
\email{luca.molinari@mi.infn.it}
\affiliation{Physics Department Aldo Pontremoli,
Universit\`a degli Studi di Milano and I.N.F.N. sezione di Milano,
Via Celoria 16, 20133 Milano, Italy.}
\date{23 Jan 2024}
%OrcId: 0000-0001-5638-8655, 0000-0002-5023-787X}
%\email{carlo.mantica@mi.infn.it, luca.molinari@unimi.it}

\begin{abstract} 
The Friedmann equations of Cotton gravity provide a simple parametrization to reproduce, by tuning a single function,
%We write the Friedmann equations for the Codazzi parametrisation of Cotton gravity, and show that they reproduce 
the Friedmann equations of several extensions of  gravity, such as $f(R)$, modified Gauss-Bonnet $f(G)$, teleparallel $f(T)$, and more. 
%cubic Einsteinian gravity and more. 
It also includes the recently proposed Conformal Killing gravity and Mimetic gravity in FRW space-times. 
%For such models the cosmological implications are the same.\\
The extensions generally have the form of a Codazzi tensor that may be associated to the dark sector.
%The general extension of the Einstein equations by a perfect fluid Codazzi tensor gives rise to a dark sector. 
Fixing it by a
suitable equation of state accomodates most of the postulated models that extend $\Lambda$CDM, as
% For example 
the Chevallier-Polarski-Lindler model.
%can be obtained without assuming dark energy as an independent physical entity.
\end{abstract}
\date{\today}
%\subjectclass[2010]{Primary: 83D05, Secondary 83C56} %Classes of solutions; algebraically special solutions, metrics with symmetries for problems in general relativity and gravitational theory 
%(Primary), 
%83C56 Dark matter and dark energy
%83C55, %Macroscopic interaction of the gravitational field with matter (hydrodynamics, etc.)
%83D05 Relativistic gravitational theories other than Einstein’s, including asymmetric field theories
%(Secondary)}
%\pacs{}
\maketitle

\section{Introduction}
In recent years there has been a flourishing of extended theories of gravity to address the problem of the Dark Sector.
They modify the Einstein equations by adding a term $H_{jk}$ to the energy-momentum tensor $T_{jk}$ of matter:
\begin{align}
R_{jk} - \frac{1}{2} g_{jk} R = \kappa (T_{jk} + H_{jk}) \label{ETFORM}
\end{align}
%$\kappa=8\pi G$.
The term originates from a new form of gravitational action or new particles.\\
Large-scale cosmology is staged in 
Friedmann-Robert\-son-Walker (FRW) space-times, where the Weyl tensor $C_{jklm}$ is zero. This fact ushers Codazzi tensors.
\medskip

%\begin{lem} 
\noindent
$\bullet $ {\em If $\nabla_m C_{jkl}{}^m=0$, then 
\begin{align}
R_{ij} - \frac{R}{2}g_{ij}=S_{ij} - g_{ij} S^k{}_k \label{SCHOUTEN}
\end{align}
where the Schouten tensor $S_{ij}=R_{ij} - \frac{1}{6} R g_{ij}$ is a Codazzi tensor, i.e.
$\nabla_i S_{jk}=\nabla_j S_{ik}$.}
\medskip
%\end{lem}

This means that $T_{jk} + H_{jk}$ has the same decomposition. The Codazzi condition 
ensures that $\nabla^k (T_{kl}+H_{kl})=0$.

Moreover, in a FRW space-time
the sum must have the perfect fluid structure of the Einstein tensor.\\
The vast majority of extended models of gravity in FRW space-times specify this property for the radiation-matter sector, with conservation.
This entails a Codazzi decomposition of the perfect fluid tensor. Then, necessarily, despite the often complex stucture of the tensor $H_{kl}$, 
the dark sector is perfect fluid and conserved.\\
%both properties separately: the matter sector
%and the dark sector are perfect fluids, and are conserved. 
%$\bullet $ {\em In a FRW spacetime the perfect fluid tensor 
%$T_{ij} = (p+\mu) u_iu_j + p g_{ij}$ is conserved $\nabla_i T^i{}_j=0$ if and only if
%$(p+\mu) u_i u_j + \frac{1}{3}\mu g_{ij}$ is a Codazzi tensor i.e. $\dot \mu = - 3H(p+\mu)$}
For these models: 
\begin{align}
&H_{jk} = \mathscr C_{jk} - g_{jk} \mathscr C^p{}_p, \label{ACCA}\\
&\nabla_i \mathscr C_{jk}=\nabla_j \mathscr C_{ik}
\end{align}

The aim of this work is to uncover this common structure, albeit the different origins of the various cosmological models.
%The inclusive paradigmatic model of all such models in FRW space-times is one wherein the perfect fluid Codazzi tensor has maximal freedom. \\
We explicitly show this in plenty of well studied extended gravity models in FRW space-times: % fall in the paradigm: 
$f(R)$, Gauss-Bonnet $f(G)$, teleparallel $f(T)$, Einsteinian cubic $f(P)$, Conformal Killing gravity, Lovelock. \\
An inclusive and simple model which they fit in, stems from Cotton gravity.

%\subsubsection*{Cotton gravity and Codazzi tensors}
In 2021 Junpei Harada \cite{Harada 21-1} introduced a modification
of General Relativity (GR) named ``Cotton gravity'' (CG), with field equations 
\begin{align}
C_{jkl}=\nabla_jT_{kl}-\nabla_kT_{jl}-\frac{1}{3}\left(g_{kl}\nabla_jT-g_{jl}\nabla_kT\right)\label{eq:Harada}
\end{align}
$T_{kl}$ is the matter energy-momentum tensor with trace $T$
and $C_{jkl}$ is the Cotton tensor:
\begin{align}
C_{jkl}\equiv \nabla_j \left (R_{kl}-\frac{R}{6}g_{kl} \right ) - \nabla_k \left (R_{jl}-\frac{R}{6} g_{jl}\right)\label{eq:eq 2}
%\nabla_j S_{kl} - \nabla_k S_{jl} \label{eq:eq 2}
\end{align}
It is $C_{jkl}=-2 \nabla_m C_{jkl}{}^{m}$. The property $g^{kl}C_{jkl}=0$ implies that $\nabla^p T_{jp}=0$.
Cotton gravity was devised so that any solution of GR is a solution of CG. 

Soon after, Harada applied his theory to describe the rotation curves of 84 galaxies
without assuming the presence of dark matter \cite{Harada 22-1}. 
The general spherically symmetric static vacuum solution 
was then obtained by Gogberashvili and Girvliani \cite{Gogberashvili 23}, with a long range
modification of Newton's law. A static solution of Cotton gravity with electric and magnetic charges 
was obtained in (\cite{Mantica23b}, eq.80)

An important progress was made in \cite{Mantica23} in showing that the equations of Cotton gravity 
are equivalent to the standard GR equations corrected by an {\em arbitrary} Codazzi tensor
\begin{align}
&R_{kl} -\frac{1}{2} Rg_{kl}=T_{kl}+\mathscr C_{kl}-g_{kl} \mathscr C^r{}_r \label{eq:Cotton Codazzi}\\
&\nabla_j\mathscr{C}_{kl}=\nabla_k\mathscr{C}_{jl} \nonumber
\end{align}
In the frame of FRW solutions, this is precisely the statement in eq.\eqref{ACCA}. 
Cotton gravity exhibits the maximal freedom in specifying $H_{kl}$. 
We refer to eq.(\ref{eq:Cotton Codazzi})
as the ``\textit{Codazzi parametrization}''.\\
While Harada's equations have third order derivatives of the metric, 
the equivalent equations \eqref{eq:Cotton Codazzi} are second order. 

%There are difficulties in finding a cosmological solution of Cotton gravity that is non-empty and conformally flat.  \\
Sussman and N\'ajera \cite{Sussman 23} used (\ref{eq:Cotton Codazzi}) to produce
FRW solutions of Cotton gravity. They posed a modified
Friedmann equation with a scalar function $\mathcal K(t)$ 
and obtained the components of a perfect fluid Codazzi tensor for which
eq.(\ref{eq:Cotton Codazzi}) are satisfied. Very recently, they published a paper \cite{SussmannB}
with several non-trivial CG solutions that generalise the well known GR solutions: FLRW, Lemaitre-Tolman-Bondi and Szekeres, as well as static perfect fluid spherically symmetric solutions (with application to galactic rotation curves) and and non-static shearfree.
%The Codazzi components used in \cite{Sussman 23} are referable to a perfect fluid.
%They first fixed a class of metrics and the structure 
%of the Ricci and stress-energy tensors. Next they obtained from eq.(\ref{eq:Cotton Codazzi})
%the components of $\mathscr{C}_{kl}$. Finally they imposed the Codazzi condition to specify
%the free parameters.\\ 
%They found a non-trivial class of FRW space-times that solve
%Cotton gravity 

Also motivated by this result, we propose a general discussion
based upon Theorem 2.1 in \cite{Mantica23}: a FRW space-time always
contains a perfect fluid Codazzi tensor that displays a freedom in its parameters. 

In Section 2 we write the Friedmann equations for Cotton gravity in a
FRW space-time.

In Section 3 we recognize that some of the most important
extended theories of gravity have the following intriguing property:
their Friedmann equations coincide with those of Cotton gravity
by a suitable choice of the Codazzi tensor. 
We show this explicitly by providing the specific Codazzi tensor for $f(R)$ gravity, Gauss-Bonnet
$f(G)$ gravity, $f(T)$ gravity, cubic Einsteinian and
$f(P)$ gravity, Lovelock.\\
These findings are well corroborated by the Generic Gravity theory by G\"urses and Heydarzade \cite{Gurses 20}, 
whose very general form of gravitational action incorporates many extended gravity theories. They show that the field equations differ from the standard FRW ones by a perfect-fluid term.

In Section 4 we show that the Codazzi parametrization
of CG extends the recently introduced conformal Killing
gravity \cite{Harada 23 a,Harada 23 b,Mantica 23 a}, at least in FRW
space-times.

%The model in \cite{Sussman 23} is a particular case.\\
In Section 5 we prove 
that the field equations of Mimetic gravity become the Cotton equations if and only if the hosting space-time
is Generalized Robertson Walker, and FRW space-times are a special case.

In Section 6 the dark sector is fixed by requesting an EoS. It 
accomodates the best known redshift dependent
models, as the Chevallier-Polarski-Lindler model.

{\em Notation}: $i,j,k,\ldots=0,1,2,3$,  $\mu,\nu, \ldots=1,2,3$. 
A dot operator $\dot X=u^k\nabla_k X$ 
is the time derivative in the comoving frame defined by $u^0=1$, $u^\mu=0$. 
$X_{[ijk]}$ is the cyclic sum $X_{ijk}+X_{kij}+X_{jki}$. 

\section{Friedmann equations of Cotton gravity in FRW space-times}

Generalized Robertson Walker space-times (GRW) are
Lorentzian manifolds that extend FRW space-times with metric 
\begin{align}
ds^2 =-dt^2 +a(t)^2 g_{\mu\nu}^\star ({\bf x}) dx^\mu dx^\nu,  \label{eq:1}
\end{align}
where $g_{\mu\nu}^\star({\bf x})$ is a positive definite metric and $a(t)$ is the scale factor. 
A covariant characterization is the existence of a vector field $u_ku^k=-1$
that is shear-free, vorticity-free and acceleration-free, and eigenvector of the Ricci tensor \cite{Mantica 17}, i.e. 
\begin{align}
&\nabla_j u_k= H(g_{jk}+u_ju_k),\label{eq:torse-forming}\\
&R_{ij}u^j =\xi u_i \label{eigenvalue}
\end{align}
where $H=\dot a/a$ is Hubble's parameter, $\xi=3(H^2+\dot H)=3\ddot a/a$.
The condition \eqref{eigenvalue} is equivalent to $\nabla_j H=-\dot H u_j$. Its 
%The Ricci tensor is
%
%\begin{align}
%R_{kl}=\frac{R-4\xi}{3}u_lu_k+\frac{R-\xi}{3}g_{kl}-2u^ju^m C_{jklm} .\label{eq:3}
%\end{align}
%
divergence and the contracted Bianchi identity give
\begin{align}
\dot R - 2\dot \xi = - 2H (R-4\xi) \label{Rdot}
\end{align}
whose solution is \cite{Capozziello22}
\begin{align}
R=%\frac{R^\star}{a^2} + 6 H^2 + 2\xi=
\frac{R^\star}{a^2} + 12H^2 + 6\dot H  \label{eq:Scalar curv}
\end{align}
where $R^{\star}$ is the spatial curvature. In $d=4$ and whenever $C_{jklm}u^m=0$ the GRW spacetime is a FRW space-time. 

In a FRW space-time the natural form of the Codazzi tensor in eq.(\ref{eq:Cotton Codazzi}) is
perfect fluid. $\Lambda g_{kl}$ with a constant $\Lambda$,
is trivially a Codazzi tensor. \\
%We choose the following perfect fluid form
This simple result is proven in \cite{Mantica23} (theorem 2.1):
\begin{prop} In a FRW space-time the tensor 
\begin{align}
\mathscr C_{kl}=\mathscr{A} u_k u_l +\mathscr{B}g_{kl}+\frac{\Lambda}{3}g_{kl}\label{eq:Codazzi Cotton}
\end{align}
is Codazzi provided that: $\nabla_j\mathscr A= - \dot{\mathscr A} u_j$,
$\nabla_j \mathscr B = - \dot{\mathscr B} u_j$, 
\begin{align}
\dot {\mathscr B} = -H \mathscr A \label{eq:condition}
\end{align}
\end{prop}
{\em Proof.}
The first two conditions mean that $\mathscr A=\mathscr A(t)$ and $\mathscr{B}=\mathscr{B}(t)$. Eq.\eqref{eq:condition} requires
$\dot{\mathscr{B}}\neq0$. 
Next, with $\nabla_j\mathscr A=-\dot{\mathscr A}u_j$,  $\nabla_j\mathscr{B}=-\dot{\mathscr{B}}u_j$,
eqs.(\ref{eq:torse-forming}) and (\ref{eq:condition}) it is:
$$ \nabla_j\mathscr C_{kl}=-u_j u_k u_l (\dot{\mathscr A}+2\dot{\mathscr B})-\dot{\mathscr B}(u_l g_{jk}+ u_k g_{jl} +u_j g_{kl} )$$
Therefore (\ref{eq:Codazzi Cotton}) is a Codazzi tensor for any choice
of the scale factor.  \hfill $\square$.

It implies that any FRW space-time is a solution of Cotton gravity with \eqref{eq:Codazzi Cotton},
and leaves an interesting degree
of freedom $\mathscr B$ in choosing the Codazzi tensor.\\
Eq.\eqref{eq:Cotton Codazzi} is written with the input \eqref{eq:Codazzi Cotton}, 
the general form of the Ricci tensor of a FRW space-time
\begin{align*}
R_{kl}=\frac{1}{3}(R-4\xi) u_lu_k+\frac{1}{3}(R-\xi)g_{kl}  %\label{eq:3}
\end{align*}
and the stress energy tensor $T_{kl}=(\mu+p)u_l u_k+pg_{kl} $ with
energy density $\mu$ and pressure $p$ of ordinary matter.\\
Contractions with 
$u^ku^l$ and $g^{kl}$ and a simple rearrangement provide the Friedmann equations
of Cotton gravity in a FRW space-time
\begin{align}
&\kappa\mu=\frac{R}{2}-\xi-3\mathscr{B}-\Lambda \label{FRIED1} \\
&\kappa p=-\frac{R}{6}-\frac{\xi}{3}+3\mathscr{B}+\frac{\dot{\mathscr{B}}}{H}+\Lambda \label{eq:Friedmann}
\end{align}
%\begin{prop} Relations \eqref{eq:Friedmann} are the Friedmann equations
%for Cotton gravity in a FRW space-time with perfect fluid
%Codazzi tensor (\ref{eq:Codazzi Cotton}).
%\end{prop}
%
They are the standard Friedmann equations of GR augmented by the Codazzi terms. Such terms naturally
correspond to the  {\sf Dark Sector}:
\begin{align}
H_{kl}=&(\mathscr{A} u_k u_l +\mathscr{B}g_{kl}) - g_{kl}(4\mathscr B -\mathscr A)\\
\equiv &(\mu_D+p_D) u_k u_l + g_{kl} p_D \nonumber
\end{align}
The function $\mathscr B(t)$ parametrizes the energy density and the pressure of the dark sector:
 \begin{align}
 &\mu_D = 3\mathscr B \label{MUD} \\
&p_D =  -3\mathscr B -\dot{\mathscr B}/H \label{PD}
 \end{align}
It implies the conservation law $\dot\mu_D =- 3H(p_D +\mu_D)$ coming from $\nabla_k H^k{}_j=0$ or the equivalent  Codazzi condition for $\mathscr{A} u_k u_l +\mathscr{B}g_{kl}$ in FRW spacetimes.

\rem{=====
by defining $\frac{R}{2}-\xi-\Lambda=-3\epsilon$. Eq.(\ref{Rdot})
ensures $\dot{\epsilon}=\frac{H}{3}(R-4\xi)$ and consequently $3\epsilon+\frac{\dot{\epsilon}}{H}=-\frac{R}{6}-\frac{\xi}{3}+\Lambda$.
The equations can be written in a more symmetric fashion by noting that the dark term is
The Friedmann equations become:
\begin{align}
&\kappa\mu=-3(\epsilon + \mathscr B )\\
&\kappa p=3 (\epsilon +\mathscr B) + \frac{1}{H} (\dot \epsilon + \dot{\mathscr B})
\end{align}
Thus $\mu=\mu_{E}+\mu_{D}$ and $p=p_{E}+p_{D}$ where the first terms
are the usual energy density and pressures in standard GR and the
second terms represent the corresponding quantities refereed to the
Dark Sector. Namely we have $\kappa\mu_{E}=-3\epsilon$, $\kappa\mu_{D}=-3\mathscr{B}$,
$\kappa p_{E}=3\epsilon+\frac{\dot{\epsilon}}{H}$and finally $\kappa p_{D}=3\mathscr{B}+\frac{\dot{\mathscr{B}}}{H}$.
It is easily verified that $\dot{\mu}_{E}+3H(\mu_{E}+p_{E})=0$ and
$\dot{\mu}_{D}+3H(\mu_D +p_D)=0$ and thus all these quantities
are conserved in Cotton gravity. 
=======}

\section{Reproducing the Friedmann equations of  extended theories}

We show that the Friedmann equations \eqref{FRIED1} and \eqref{eq:Friedmann} of Cotton gravity
may reproduce the Friedmann equations of other extended theories in
absence of cosmological constant. 
With eq.(\ref{eq:Scalar curv}) and $\xi=3(H^2+\dot H)$ we write
them as 
\begin{align}
& \kappa \mu = \frac{R^\star}{2a^2} + 3 H^2 - 3 \mathscr B  \label{CGFRIEDMANN1}\\
& \kappa p= - \frac{R^{\star}}{6a^2} - 3H^2 -2\dot H + 3\mathscr B + \frac{\dot{\mathscr B}}{H} \label{eq:Friedman def}
\end{align}
The comparison with the Friedmann equations of other theories 
selects the function $\mathscr{B}(t)$ that reproduces them. %
%

%\section{Generic gravity}

In ref.\cite{Gurses 20} G\"urses and Heydarzade introduced the Generic Gravity Theory. It is characterized by a very general form of gravitational action, with 
a scalar function $\mathcal F$ of the metric tensor, the Riemann tensor and its covariant derivatives at any order:
\begin{align}
S=& \int d^4x \sqrt{-g} \Big[\frac{R-2\Lambda}{\kappa} \nonumber\\
&+\mathcal F (g,\, {\rm Riem},\, \nabla {\rm Riem},\nabla\nabla \rm {Riem},\, ...)\Big]+S_{mat}
\end{align}
The theory contains all modified theories of gravity
based on curvature such as $f(R)$, $f(G)$, $f(P)$ theories. \\
The authors prove a theorem for Generic Gravity in 
FLRW cosmology (\cite{Gurses 20}, Theorem 5): the field equations always
take the form $G_{kl}=\kappa T_{kl}+H_{kl}$, where
$H_{kl}=\mathbb{A}g_{kl}+\mathbb{B}u_{k}u_{l}$ accounts for the contribution of all aforementioned higher
order terms. The explicit expressions for $\mathbb A$ and $\mathbb B$ was given for the Einstein-Lovelock and for generalized Einstein-Gauss-Bonnet theories. In \cite{Gurses 24} the explicit analysis is extended to quadratic gravity. 

In this general setting, we note the following 
\begin{lem}
In a FLRW space-time if $H_{kl}=\mathbb{A}g_{kl}+\mathbb{B}u_{k}u_{l}$
is divergence-free then it is 
$$H_{kl}=\mathscr{C}_{kl}-g_{kl}\mathscr{C}_{p}^{p}$$
being $\mathscr{C}_{kl}$ a Codazzi tensor.

Proof.  The divergence-free condition is $3H\mathbb{B}=\dot{\mathbb{A}}-\dot{\mathbb{B}}.$
Let $\mathscr{A}=\mathbb{B}$ and $\mathscr{B}=\frac{1}{3}(\mathbb{B}-\mathbb{A})$, 
then  $\dot{\mathscr{B}}=-H\mathscr{A}$. By 
{\rm Proposition 1} the tensor $\mathscr{C}_{kl}\equiv\mathscr{A}u_{k}u_{l}+\mathscr{B}g_{kl}$
satisfies the Codazzi condition. \hfill $\square $
\end{lem}
{\sf Thus we may state that in FLRW cosmology Cotton Gravity is
equivalent to any Generic Gravity theory. }

\subsection{$\mathbf{f(R)}$ gravity}
Perhaps it is the best known extended theory of gravity. It was
introduced by Buchdahl in 1970 \cite{Buchdahl 70} and gained popularity 
with the works on cosmic inflation by Starobinsky \cite{Starobinsky 80}. 
Recently $f(R)$ theories are possible candidates
to explain the observed cosmic acceleration.\\
Investigations to explain both dark energy and inflation
were pursued in the papers by Cognola et al. \cite{Cognola 08}, Nojiri and Odintsov \cite{Nojiri 03,Nojiri 07}.
Capozziello considered $f(R)$  to discuss
the issue of quintessence \cite{Capozziello 02}. 
For general reviews on $F(R)$ see \cite{Nojiri 17,Nojiri 11,Sotiriu10}.

The action of $f(R)$ gravity is
\begin{align*}
S=\frac{1}{2\kappa} \int d^{4}x\sqrt{-g} f(R) + S^{(m)}
\end{align*}
where $S^{(m)}$ is the matter term. With $f_R =df/dR$, the field
equations are \cite{Sotiriu10}:
\begin{align}
f_R R_{kl} - \frac{f}{2} g_{kl} - (\nabla_k\nabla_l - g_{kl}\square ) f_R = \kappa T_{kl}\label{eq: f(R)}
\end{align}
They can be rewritten in the form \eqref{ETFORM}. In \cite{Capozziello 19} it was proven that in a FRW space-time
the resulting term $H_{jk}$ is a perfect fluid tensor. \\
For the spatially flat ($R^\star=0)$ FRW space-time the Friedmann equations
of $f(R)$ gravity are (\cite{Sotiriu10}, eqs.75, 76):
\begin{align}
& \kappa \mu = 3 f_R H^2 - \tfrac{1}{2}(R f_R -f) + 3H \dot R f_{RR}  \label{FR1}\\
&(3H^2+2\dot H)f_R = -\big [\kappa p+ \dot R^2 f_{RRR}+2H \dot R f_{RR} \label{FR2} \\
&\qquad\qquad\qquad\qquad +\ddot R f_{RR}+\tfrac{1}{2}(f-R f_R) \big ] \nonumber
\end{align}
In comparing \eqref{FR1} with \eqref{CGFRIEDMANN1} we identify 
\begin{align}
\mathscr{B}=H^{2}(1-f_R)
+\tfrac{1}{6}(Rf_R -f) -H\dot{R}f_{RR} \label{eq:B For f(R)}
\end{align}
In computing $\dot{\mathscr{B}}$ we note that $\dot f(R)=f_R(R)\dot{R}$,
$\dot{f_R}(R)=f_{RR} \dot R$, $\dot f_{RR} =f_{RRR}\dot R$, so that
\begin{align*}
\dot{\mathscr B}= 2H\dot{H}(1-f_R)+f_{RR}(\frac{R}{6}- \dot H-H^2)\dot R\\
-H\ddot{R}f_{RR}-H\dot R^2 f_{RRR}
\end{align*}
The restriction $R^{\star}=0$ in \eqref{eq:Scalar curv} gives $R=12H^{2}+6\dot{H}$.\\ 
We obtain $ - \mathscr A $:
\begin{align}
\frac{\dot{\mathscr B}}{H}= 2 \dot H (1-f_R) + (H \dot R - \ddot R) f_{RR} 
- \dot R^2 f_{RRR}  \label{eq:A for f(R)}
\end{align}
Using now \eqref{eq:Friedman def} we obtain \eqref{FR2}.

\begin{prop}\label{6} The Friedmann equations of Cotton gravity with the perfect
fluid Codazzi tensor \eqref{eq:Codazzi Cotton} are the Friedmann
equations of $f(R)$ gravity with the choice \eqref{eq:B For f(R)}. %and \eqref{eq:A for f(R)}.
\end{prop}
\subsection{$\mathbf{f(G)}$ gravity}
A second well known extended theory that tries to solve the problem
of dark energy is the Gauss-Bonnet gravity, alias 
$f(G)$ gravity  \cite{Nojiri 05,Nojiri 06,Cognola 06}.
\begin{align*}
S=\int d^4 x \sqrt{-g}\left[\frac{R}{2\kappa}+f(G)\right]+S^{(m)}
\end{align*}
where $ G =R^2-4R_{kl} R^{kl}+R_{jklm}R^{jklm} $ is the Gauss-Bonnet invariant.\\
The field equations may be written in the form $R_{kl}-\frac{1}{2} R g_{kl}=\kappa (T_{kl}+H_{kl})$
with the following divergence-free tensor $H_{kl}$ \cite{Capozziello 19 a, Gurses 20}: 
\begin{align}
H_{kl}&=\tfrac{1}{2} g_{kl} f - 2 f_G (R R_{kl} - 4 R_{kq} R_l^{q} + 2 R_k\,^{pqr} R_{lpqr}) \nonumber\\
& -  4 f_G R_k\,^{pq}\,_lR_{pq}+2R (\nabla_k\nabla_l f_G -g_{kl} \square f_G ) \nonumber \\
&-4(R_l{}^p \nabla_p \nabla_k f_G + R_k{}^p \nabla_p\nabla_l f_G) +4(\square f_G )R_{kl}\nonumber\\
&+4 (R^{pq}g_{kl} -R_k\,^{pq}\,_l) \nabla_p \nabla_q f_G \label{eq:gauss Bonnet}
\end{align}
where $f_G=d f/d G$. In a FRW space-time it is a perfect fluid tensor.
For the spatially flat case,  the Gauss-Bonnet invariant is $G=24(\dot H H^2+H^4)$
and the Friedmann equations
of $f(G)$ gravity are expressible as (eq.5 in \cite{Myrzakulov 11})
\begin{align}
&\kappa\mu = 3H^2 - \kappa(G f_G - f - 24H^3 \dot G f_{GG} )\\
&\kappa p= - 3 H^2 -2 \dot H + \kappa( G f_G - f) \label{eq:Friedmann for Gaus Bonnet} \\
&\qquad -16 \kappa H  (H+\dot H) f_G - 8\kappa H^2 \ddot f_G  \nonumber
\end{align}
With $\dot f = f_G \dot G$ we obtain $\dot f_G =f_{GG} \dot G$
and the first equation rewrites as 
$\kappa\mu = 3 H^2 -\kappa(G f_G - f - 24 H^3 \dot f_G )$.\\
Comparison with \eqref{CGFRIEDMANN1} gives the following
identification 
\begin{align}
\mathscr{B}=\kappa(G f_G - f - 24 H^3 \dot f_G)\label{eq:B for Gauss Bonnet}
\end{align}
After straightforward calculations we infer 
\begin{align}
-\mathscr A=\frac{\dot{\mathscr B}}{H}=8 \kappa [\dot f_G ( H^3 - 2H \dot H )-H^2 \ddot f_G ].   \label{eq:A for Gauss bonnet}
\end{align}
Thus from \eqref{eq:Friedman def} we obtain eq.(\ref{eq:Friedmann for Gaus Bonnet}).
\begin{prop}\label{7} The Friedmann equations of Cotton gravity with the perfect
fluid Codazzi tensor \eqref{eq:Codazzi Cotton} are the Friedmann equations of Gauss-Bonnet $f(G)$ gravity 
with the choice \eqref{eq:B for Gauss Bonnet}. %and \eqref{eq: B for F(T)}.
\end{prop}
\subsection{$\mathbf{f(T)}$ gravity}
In the framework of gravity theories with torsion, the ``teleparallel
equivalent of general relativity'' is the best known one.
It is widely discussed in \cite{Cai16} and briefly reviewed in
\cite{Harko 14}. It is based on the action
(\cite{Harko 14} eq. 2.5):
\begin{align*}
S=\frac{1}{2\kappa}\int d^{4}x\sqrt{-g}\left[T+f(T) \right] +S^{(m)}
\end{align*}
The field equations are  eqs.~263 in \cite{Cai16} or eq.~2.6 in \cite{Harko 14}. For the spatially flat FRW space-time, 
the Friedmann equations of $f(T)$ gravity are expressible as (eqs 2.9, 2.10 in \cite{Harko 14} or  eqs.~267, 268 in \cite{Cai16}):
\begin{align}
&H^2 =\frac{\kappa}{3}\mu-\frac{f}{6}-2f_T H^{2} \label{fT1}\\
&\dot H=- \frac{1}{2}\frac{\kappa (p+\mu)}{1+f_T - 12 H^2 f_{TT}}
\label{fT2}
\end{align}
The torsion scalar is $T=-6H^{2}$ (eqs.269 in \cite{Cai16}), and $f_T=df/dT$. 
Comparing with eq.\eqref{CGFRIEDMANN1}
we identify 
\begin{align}
\mathscr{B}= - \frac{1}{6} f(T) - 2 f_T (T) H^2  \label{eq: B for F(T)}
\end{align}
Note that $\dot{f}(T)=f_T\dot T=-12H\dot H f_T$,
$\dot f_T = f_{TT} \dot T=-12H\dot H f_{TT}$. Thus 
$\dot{\mathscr B}=-2H\dot H f_T +24 H^3 \dot H f_{TT}$ and
\begin{align}
- \mathscr A =\frac{\dot{\mathscr B}}{H}=-2\dot H f_T (T)+ 24 H^2 \dot H f_{TT}(T)\label{eq A for F(T)}
\end{align}
Summing the Friedmann equations \eqref{CGFRIEDMANN1} and \eqref{eq:Friedman def} of Cotton gravity 
we  get
\begin{align}
\frac{\kappa}{2} (p+\mu) = -\dot H +\frac{\dot{\mathscr B}}{2H}      \label{eq:21}
\end{align}
Inserting (\ref{eq A for F(T)}) in (\ref{eq:21}) gives eq.\eqref{fT2}.
%\begin{align*}
%\frac{\kappa}{2}(p+\mu)=-\dot H (1+f_T -12 H^2 f_{TT} )
%\end{align*}
%
\begin{prop}\label{8} The Friedmann equations of Cotton gravity with the perfect
fluid Codazzi tensor \eqref{eq:Codazzi Cotton} are the Friedmann equations of $f(T)$ gravity with the
choice \eqref{eq: B for F(T)}. % and \eqref{eq A for F(T)}.
\end{prop}

\subsection{Einsteinian Cubic and $\mathbf{f(P)}$ gravity}
In \cite{Bueno} an extended theory is proposed, based on
an invariant $P$ constructed with cubic contractions of
the Riemann tensor. The theory was subjected to three constraints:
1) the spectrum
%\footnote{the physical modes that propagate in the linearized metric around a maximally symmetric background.} 
should be identical to that of GR (whence the
name); 2) it is neither topological nor trivial
in d=4; 3) it is independent of the dimension.\\ 
The action is 
\begin{align}
&S=\int d^{4}x\sqrt{-g}\left[\frac{R-2\Lambda}{2\kappa}+ P \right] +S^{(m)} \nonumber\\
&P= -\beta_1 R_j\,^{pq}\,_k R_p\,^{rs}\,_q R^j\,_{rs}\,^k+\beta_{2}R_{jk}\,^{rs}R_{rs}\,^{pq}R_{pq}\,^{jk}\nonumber \\
&+\beta_{3}R^j{}_k R_{pqrj}R^{pqrk} +\beta_4 R_{pqrs}R^{pqrs} + \beta_5  R_{jkpq}R^{kp}R^{jq} \nonumber\\
& + \beta_6 R_k^{p}R_{p}^j R_j^k+\beta_7 R R_{pq} R^{pq}+\beta_8 R^3 
\end{align}
The aforementioned constraints impose three linear relations
among the coefficients $\beta_i$.\\ 
In \cite{Erices 19} the cosmological applications of Einsteinian cubic gravity
at early and late times were investigated. In \cite{Marciu 23} 
the viability of the theoretical model is analyzed, by considering 
observational features such as cosmic chronometers data,
baryon acoustic oscillations, and supernovae.

The field equation may be written in the  form 
$R_{kl}-\frac{1}{2}Rg_{kl}+\Lambda g_{kl}=\kappa(T_{kl}+ H_{kl})$
where $H_{kl}$ is an involved symmetric tensor containing contractions
of the Riemann and the Ricci tensor. \\
The Friedmann equations are eqs.11 and 12 in \cite{Erices 19}:
\begin{align}
&3H^2=\kappa ( \mu+6\alpha\widetilde{\beta} H^6)+\Lambda\\
&3H^2+2\dot H = - \kappa\left [ p - 6\alpha\widetilde{\beta}H^4 (H^2+2 \dot H)\right ]+\Lambda
\label{eq:Friedmann for Cubic}
\end{align}
with $\widetilde{\beta} = -\beta_1 + 4\beta_2 +2\beta_3 + 8\beta_4$. Comparison with
\eqref{CGFRIEDMANN1} yields
\begin{align}
\mathscr{B}=6\kappa\widetilde{\beta}H^6  \label{eq:B for cubica}
\end{align}
Then
%
%\begin{align}
$ - \mathscr A = \dot{\mathscr B}/H=12 \kappa \widetilde{\beta} H^4 \dot H $. % \label{eq:A for cubic}
%\end{align}
From eq.(\ref{eq:Friedman def}) we get eq.(\ref{eq:Friedmann for Cubic}). 
\begin{prop} The Friedmann equations of  Cotton gravity with the perfect
fluid Codazzi tensor \eqref{eq:Codazzi Cotton} are the Friedmann 
equations of cubic Einsteinian gravity
with the choice \eqref{eq:B for cubica}. % and \eqref{eq:A for cubic}.}
\end{prop}

In the same paper \cite{Erices 19} the authors proposed the $f(P)$ extension
of Einsteinian cubic gravity:
\begin{align*}
S=\int d^{4}x\sqrt{-g}\left[\frac{R}{2\kappa}+f(P)\right] +S^{(m)}
\end{align*}
The field equations are still of the type $R_{kl}-\frac{1}{2}Rg_{kl}=\kappa(T_{kl}+\widetilde{H}_{kl})$
where $\widetilde{H}_{kl}$ is quite more involved. The
 Friedmann equations are eqs. 26, 27
in \cite{Erices 19}:
\begin{align}
&3H^2 = \kappa\mu- \kappa f-18 \kappa \widetilde{\beta} H^4 (H\partial_t - H^2 -\dot H) f_P \label{FP1}\\
&3H^2+2\dot H = -\kappa P-\kappa f - 6 \kappa \alpha\widetilde{\beta} H^3  \label{fP2}\\
& \qquad \times \left[H\partial_t^2 +2(H^2+2\dot H)\partial_t  -3H^3 -5H\dot H \right  ] f_P\nonumber
\end{align}
where $f_P=d f/d P$ and $P=6\widetilde{\beta}H^{4}(H^{2}+2\dot{H})$.\\
It is simple to identify
\begin{align}
\mathscr B= - \kappa f - 18\kappa\widetilde{\beta}H^4 (H \dot{f_P} - H^2 f_P - \dot H f_P)\label{eq:B for f(P)}
\end{align}
Now $\dot f(P)=f_P\dot P = 18 \widetilde{\beta} H^3 (4H^2 \dot H +4\dot H^2 + H\ddot H )f_P$.
After tedious but straightforward calculations it is %$- \mathscr A= $
\begin{align}
 \frac{\dot{\mathscr B}}{H}=&6 \kappa\widetilde{\beta} H^3 [2 H \dot H f_P 
- \dot f_P ( 4\dot H - H^2) - H \ddot f_P ] \label{eq: A for f(P)} 
\end{align}
Thus from (\ref{eq:Friedman def}) we get \eqref{fP2}.
\begin{prop}\label{10} The Friedmann equations of Cotton gravity with the perfect
fluid Codazzi tensor (\ref{eq:Codazzi Cotton}) are the Friedmann
equations of $f(P)$ gravity with
the choice \eqref{eq:B for f(P)}.% and \eqref{eq: A for f(P)}.
\end{prop}

\subsection{Regularized cubic Lovelock gravity}% in four dimensions}

In Section 3 of \cite{Casalino 21} the authors focused on the cubic Lovelock gravity in a 4-dimensional 
FRW space-time. They obtained
the following Friedmann equations 
\begin{align}
& \kappa\mu=3J^2(1+\alpha J^2+\beta J^4) \label{FLOV1}\\
& \kappa(p+\mu)=- 2(\dot H -\frac{R^\star}{6a^2})(1+2\alpha J^2 + 3\beta J^4)
\label{eq:Friedmann Lovelock}
\end{align}
where $J^2= H^2+\frac{R^\star}{6a^2}$. We dropped their cosmological constant and
the stress energy tensor is multiplied by a factor 2 to match 
our notation. We state the following:
\begin{lem}
If in (\ref{eq:Friedman def}) we put $-\mathscr{B}=F(J^{2})$
where $F$ is a smooth arbitrary function of $J^{2}=H^{2}+\frac{R^\star}{6a^{2}}$
then 
\begin{equation}
\frac{\kappa}{2}(p+\mu)=-(\dot{H}-\frac{R^{\star}}{6a^{2}})[1+F_J (J^2)]
\end{equation}
where $F_J=dF/dJ^2$.\\
{\bf Proof.}
From \textit{(\ref{eq:Friedman def}}) we have $\frac{\kappa}{2}(p+\mu)=\frac{R^{\star}}{6a^{2}}-\dot{H}+\frac{\dot{\mathscr{B}}}{2H}$.
If\textit{ $-\mathscr{B}=F(J^{2})$ }then $\dot{\mathscr{B}}=-2F_{J^{2}} J\dot J$.
On the other hand $J\dot{J}=H(\dot{H}-\frac{R^{\star}}{6a^{2}})$ so
that $\frac{\dot{\mathscr{B}}}{2H}=-(\dot{H}-\frac{R^{\star}}{6a^{2}})F_{J^{2}}$
and the Lemma is proven. \hfill $\square$
\end{lem}
Choose $F_J (J^{2})=\alpha J^{2}+\beta J^{4}$ and \eqref{FLOV1}\eqref{eq:Friedmann Lovelock}
are recovered.

\subsection{Sussman N\'ajera Model in Cotton gravity}
In  \cite{Sussman 23,SussmannB} the authors introduced the following modified
Friedmann equation:
\begin{align}
H^2 = \frac{\kappa}{3} \mu - \frac{R^\star}{6a^2} - \frac{\gamma}{a^2} \mathcal K (t)\label{eq:Friedmann for suss}
\end{align}
with an arbitrary dimension-less function $\mathcal K(t)$ and a constant $\gamma$. Then they computed the components of the Codazzi tensor that solves Cotton gravity.
Comparison of \eqref{eq:Friedmann for suss} with \eqref{CGFRIEDMANN1} gives $\mathscr B$ whence $\mathscr A$ is
computed:
\begin{align}
&\mathscr B = - \frac{\gamma\mathcal K(t)}{a^2} \label{eq: B for Sussman}\\
&\mathscr A =-\frac{\dot{\mathscr B}}{H} =  \frac{\gamma\dot{\mathcal K}(t)}{a^2 H} - \frac{2\gamma\mathcal K(t)}{a^2}  \label{eq:A for Sussman}
\end{align}
The expression of the Cotton tensor 
\begin{align}
\mathscr C_{kl} = \left [ \frac{\gamma{\dot{\mathcal K}} (t)}{a^2 H} - \frac{2\gamma\mathcal K(t)}{a^2} \right ] u_k u_l 
- \frac{\gamma\mathcal K(t)}{a^2} g_{kl} \label{eq: Codazzi for Sussman}
\end{align}
compares with the components in eq.19 evaluated in \cite{SussmannB}.

%

%Thus from eq.(\ref{eq:Friedman def}) we get 
%
%\begin{align}
%\kappa p= - \frac{R^\star}{6a^2} -3H^2 - 2\dot H - \frac{\gamma\dot{\mathcal K (t)}}{a^2 H}-\frac{\gamma\mathcal K(t)}{a^2}\label{eq:secon fried Suss}
%\end{align}

\section{Comparison with conformal Killing gravity}
After Cotton gravity, Harada introduced a new theory of gravity to explain the present accelerated phase
of the Universe without explicit introduction of dark energy \cite{Harada 23 a}:
%It is defined by the field equations 
%
\begin{align}
\nabla_{[j} R_{kl]} -\frac{1}{3}\nabla_{[ j } R g_{kl]} = \nabla_{[j}T_{kl]} -\frac{1}{6} \nabla_{[j} T g_{kl]} \label{eq:Killing Harada}
\end{align}
The equations are manifestly of third order in the derivatives
of the metric tensor.\\
Shortly after in \cite{Mantica 23 a} we introduced a parametrization of the theory by showing that
(\ref{eq:Killing Harada}) is equivalent to the Einstein's equation in which the stress-energy
tensor is augmented by a divergence-free conformal Killing tensor:
\begin{align}
&R_{kl}-\frac{1}{2}Rg_{kl}=T_{kl}+K_{kl} \label{eq:Conformal KIlling mantica}\\
&\nabla_{[j}K_{kl]}  =\frac{1}{6}\nabla_{[j}K g_{kl]} \label{CKTDIV}
\end{align}
where $K=g^{pq}K_{pq}$. We named this theory {\em conformal Killing gravity}.\\
The second equation
defines a divergence-free conformal Killing tensor (CKT). They are deeply
investigated in differential geometry and in physics \cite{Coll 06,kobialko 22,Rani 03,Step12}.

We proved existence of a conformal Killing tensor in any FRW space-time, 
obtaining two modified Friedmann equations that
allow for the presence of a dark sector. When applied to a simple
toy model, this theory reveals a phantom dark fluid with equation
of state (EoS) parameter $w=-5/3$ \cite{Mantica 23 a}.\\
In a second paper \cite{Harada 23 b}
Harada developed an interesting cosmological analysis confirming in
general that the dark energy predicted by the conformal Killing gravity
has the same EoS parameter.\\ 
Here we investigate the connections between Cotton and Conformal Killing gravity.

To this end, consider a generic space-time endowed
with a (0,2) symmetric tensor satisfying the relation
\begin{align}
\nabla_j K_{kl}= a_j g_{kl} - b_k g_{jl} - b_l g_{jk} \label{eq:Sinyukov}
\end{align}
We call such tensors Sinyukov-like (see \cite{Mantica12},\cite{Step12}). 
If $K_{kl}$ is the Ricci tensor we recover the Sinyukov manifolds, investigated for example in \cite{Formella 95}.

Here we consider divergence-free Sinyukov-like tensors:
\begin{align}
\nabla_j K_{kl}=5\frac{\nabla_jK}{18}g_{kl}-\frac{\nabla_kK}{18}g_{jl}-\frac{\nabla_lK}{18}g_{jk}\label{Syniukov div}
\end{align}
where $K=K^p{}_p$. They satisfy the condition \eqref{CKTDIV} that defines divergence-free conformal
Killing tensors (CKT)  \cite{Rani 03}. 
%
%\begin{align}
%\nabla_{[j} K_{kl]} =\frac{1}{6} \nabla_{[j} K g_{kl]} \label{eq:Conforma Killing tensor}
%\end{align}
%

A space-time with a Sinyukov-like divergence-free tensor
is a solution of conformal Killing gravity (\ref{eq:Conformal KIlling mantica}).\\
On the other hand (\ref{Syniukov div}) implies the Codazzi condition
\begin{align}
\nabla_j \left [K_{kl}-\frac{K}{3} g_{kl}\right  ] =\nabla_k \left [ K_{jl}-\frac{K}{3} g_{jl} \right ]
\end{align}
Then $\mathscr C_{kl}=K_{kl}-\frac{1}{3}Kg_{kl}$ is
a Codazzi tensor with $\mathscr{C}_{r}^{r}=-\frac{1}{3}K$. From $R_{kl}-\frac{1}{2}Rg_{kl}=T_{kl}+K_{kl}$
we recover the paradigm \eqref{eq:Cotton Codazzi}
%$R_{kl}-\frac{1}{2} R g_{kl} = T_{kl}+\mathscr C_{kl} - g_{kl} \mathscr C_r^r $
%and the Cotton field equations (\ref{eq:Cotton Codazzi}) are satisfied
with the same stress-energy tensor. 
\begin{prop}\label{3} A space-time with a divergence-free Sinyukov-like
tensor (\ref{Syniukov div}) is a solution both of conformal Killing gravity
(\ref{eq:Conformal KIlling mantica}) and of Cotton gravity (\ref{eq:Cotton Codazzi}), 
with the same stress-energy tensor.
\end{prop}

We show that, rather surprisingly, any FRW space-time is equipped with a Sinyukov-like
tensor.\\
We recall that a vector $Z_j$ is a conformal
Killing vector \cite{Rani 03} (CKV for short) if the following condition
holds:
\begin{align}
\nabla_jZ_{i}+\nabla_{i}Z_j=2\psi g_{ij}\label{eq:conformal vect}
\end{align}
where the scalar function $\psi$ is called conformal factor. 

Let $Z_j=Fu_j$ with $u_ju^j=-1$ and $F$ a scalar function. The following result holds: \\
{\em In a GRW space-time $Fu_j$
is a CKV if and only if $\dot F=HF=\psi $ and  $\nabla_{i}F=-u_{i}\dot{F}$, i.e.
$F$ depends only on time.} (\cite{Shenawy 07} thrm 2.1,  \cite{Maartens 86} thrm 1)

In this case, since $H=\dot a/a$, we obtain $F(t)=ka(t)$ for some constant $k$. 
According to Rani et al. \cite{Rani 03}, the CKV originates a conformal Killing tensor 
\begin{align}
K_{ij}= F^2 u_i u_j+ F_1 g_{ij}\label{eq:eq10}
\end{align}
for arbitrary scalar function $F_{1}$. Let us choose $F_1$ in order that 
$0=\nabla_p K^p{}_j$. A simple evaluation using (\ref{eq:torse-forming}) shows that $\nabla_i F_1= - 5F\dot F u_i$.
Then $F_1$ depends only on time, and $\dot F_1= 5F\dot F$. An integration gives
$F_1 = \frac{5}{2} F^2 - \Lambda$ being $\Lambda $ a constant. Now 
\begin{align}
K_{jk} =F^2 (u_ju_k + \frac{5}{2} g_{jk}) -\Lambda g_{jk} \label{CKT30}
\end{align}
Next evaluate
%The evaluation of the gradient gives:
$\nabla_i K_{jk} = H F^2 (-5u_i g_{jk}+ g_{ij}u_k + g_{ik}u_j)$. Contraction with $g^{jk}$: $\nabla_i K = -18 HF^2 u_i$. It turns out that
$K_{jk}$ satisfies \eqref{Syniukov div}, i.e. it is divergence-free Sinyukov-like. \hfill $\square$  \\
The associated Codazzi tensor $\mathscr C_{kl}=K_{kl}-\frac{K}{3}g_{kl}$ is 
\begin{align}
\mathscr C_{ij}= F^2 (u_i u_j -\frac{1}{2}g_{ij}) + \frac{\Lambda}{3}  g_{ij} \label{eq:Codazzi Sinyukov}
\end{align}
We have proven
\begin{prop}\label{5} Any GRW space time, and thus any FRW
space-time, is a solution of both Cotton and conformal Killing gravity
with the same stress-energy tensor.
\end{prop}

The Codazzi tensor \eqref{eq:Codazzi Sinyukov} is not as general as \eqref{eq:Codazzi Cotton},
since it is fixed up to a constant. In fact the condition (\ref{eq:condition})
is more general than $\dot F=HF$. In a FRW space-time 
Cotton gravity is more general than conformal Killing gravity.

The Conformal Killing tensor \eqref{CKT30} is used in \cite{Mantica 23 a} to obtain the
Friedmann equations of conformal Killing gravity. The eigenvalue equation
$K_{ij}u^i=\lambda u_j$ gives $\lambda=\frac{3}{2} F^{2}-\Lambda$
and thus $K=6\lambda+2\Lambda$. We rewrite the tensor as
\begin{align}
K_{ij}= %&\frac{K-4\lambda}{3}u_{i}u_j+\frac{K-\lambda}{3}g_{ij}\\
\frac{2\lambda+2\Lambda}{3}u_{i}u_j+\frac{5\lambda+2\Lambda}{3}g_{ij}\label{eq:eq 20}
\end{align}
Note that $2\lambda  = 3 F^2 -2\Lambda = 3k^2 a^2(t) - 2\Lambda$.
%
%\begin{align}
%\frac{\dot \lambda}{2\lambda+\Lambda} = \frac{\dot F}{F} =H=\frac{\dot a}{a}
%\end{align}
%From this we get $2\lambda+\Lambda=Ca^{2}$ being $C$ a suitable
%constant. Thus we recover the results in \cite{Mantica 23 a}.

\section{Comparison with Mimetic gravity}

In 2013, Chamseddine and Mukhanov \cite{Chamseddine 13,Chamseddine 14,Chamseddine 17}
proposed a modification of GR where the conformal degree of freedom is distinguished.
This is done by parametrizing the physical metric tensor $g_{kl}$ in terms 
of an auxiliary metric $\widetilde{g}_{kl}$ and a scalar field $\phi$, called {\em mimetic field}:
\begin{equation}
g_{kl}(\widetilde{g},\phi)= - (\widetilde{g}^{pq}\nabla_{p}\phi\nabla_{q}\phi) \,\widetilde{g}_{kl}
\label{eq:Mimetic metric}
\end{equation}
where $\widetilde g^{pq} \equiv (\widetilde g^{-1})_{pq}$. Then $g^{kl} = - (\widetilde{g}^{pq}\nabla_{p}\phi\nabla_{q}\phi)^{-1}\,\widetilde{g}^{kl}$. The compatibility condition follows:
\begin{align}
g^{kl}\nabla_k \phi \nabla_l \phi  = - 1 \label{compat}
\end{align}
A conformal transformation of the auxiliary metric $\widetilde{g}'_{kl}=\Omega^{2}\widetilde{g}_{kl}$
leaves the physical metric invariant. Mimetic gravity may be viewed as a conformal extension
of Einstein theory, which is locally Weyl invariant:
this fact was pointed out by Barvinsky \cite{Barvinsky 14}.

The gravitational action depends upon the auxiliary
metric and the mimetic field. Alternatively, it depends on the physical metric but 
with the constraint \eqref{compat}:
\begin{equation}
S=\int d^{4}x\sqrt{-g}\left[R+\zeta(g^{pq}\nabla_{p}\phi\nabla_{q}\phi+1)-V(\phi)\right ]+S^{(m)}\label{eq:Lagrang mimetic}
\end{equation}
$V(\phi)$ is a potential and $\zeta $ is a Lagrange multiplier. %depending on the mimetic field 
%and $\zeta$ is the Lagrange multiplier.\\
For a thorough review of Mimetic gravity see \cite{Sebastiani 17}.\\ 
The first field equation is obtained by minimising with respect to the metric: 
\begin{equation}
R_{kl}-\frac{1}{2}Rg_{kl}=T_{kl}+2\zeta\nabla_{k}\phi\nabla_{l}\phi+g_{kl}V(\phi)\label{eq: Mimetic equations}
\end{equation}
It has the form of an extended theory with dark sector
explicitly represented by the mimetic field (whence the name of `mimetic dark matter' in the literature).\\ 
The trace and the constraint give $2\zeta=R+T+4V$. The covariant divergence of (\ref{eq: Mimetic equations}) is 
\begin{equation}
2[\nabla^{k}\zeta \nabla_{k}\phi +\zeta \nabla^k\nabla_k\phi]\nabla_{l}\phi +\nabla_{l}V=0\label{eq:conservation mimetic}
\end{equation}
where we used $\nabla_kR^k{}_l-\frac{1}{2}\nabla_l R=0$, $\nabla^{k}T_{kl}=0$
and $\nabla_{j}(\nabla^{p}\phi \nabla_{p}\phi)=0$.

Variation of the action with respect to the mimetic field gives 
\begin{equation}
2\nabla^{p}(\zeta\nabla_{p}\phi)=-\frac{\partial V}{\partial\phi}
\end{equation}
Since $g^{pq}\nabla_{p}\phi\nabla_{q}\phi=-1$ the vector field
$u_{k}=-\nabla_{k}\phi$ is unit time-like and closed, i.e. $\nabla_{j}$$u_{k}=\nabla_{k}u_{j}$.
Then it is vorticity-free and acceleration-free:
\begin{equation}
\nabla_{j}u_{k}=H(g_{jk}+u_{j}u_{k})+\sigma_{jk}\label{eq: torse forming shear}
\end{equation}
being $\sigma_{jk}$ the shear tensor. The corresponding metric is
 (see \cite{Chamseddine 17,Chamseddine 13} and \cite{Coley 94})
\begin{equation}
ds^{2}=-dt^{2}+g_{\mu\nu}^{\star}({\bf x},t)dx^{\mu}dx^{\nu}\label{eq: metric shear}
\end{equation}
By fixing the hypersurfaces of constant
time of (\ref{eq: metric shear}) to be of constant $\phi$, the solution of the
constraint $g^{pq}\nabla_{p}\phi\nabla_{q}\phi=-1$ may be written 
(see \cite{Chamseddine 13} and reference therein or \cite{Chamseddine 17}):
\begin{equation}
\phi=\pm t+const.\label{eq:phi}
\end{equation}
Thereby choosing $\phi=t $ and using $u_{k}=-\nabla_{k}\phi$,  it is $u_{0}=-1,u_{\mu}=0$. \\
We then conclude that the general metric for mimetic gravity is (\ref{eq: metric shear}).
In this context $V=V(t)$, while in general $\zeta $ is a function of ${\bf x}$ and $t$.\\
The field equations take the form 
\begin{equation}
R_{kl}-\frac{1}{2}Rg_{kl}=T_{kl}+2\zeta u_{k}u_{l}+V g_{kl} \label{eq:Mimetic final}
\end{equation}
and (\ref{eq:conservation mimetic}) becomes
%
%\begin{equation}
$\nabla_{l}V=-2( \dot{\zeta}+3 H\zeta) u_{l} $. %\label{eq: cons mimetic V}
%\end{equation}
Transvecting it with $u^{l}$ gives the interesting relation
\begin{equation}
\dot{V}=2\dot{\zeta}+6H\zeta\label{eq: cons mimetic final}
\end{equation}
where we used $\nabla_{p}u^{p}=3H$ derived from (\ref{eq: torse forming shear}).\\
Now note that (\ref{eq:Mimetic final}) 
%has the most similar
%appearance with Cotton gravity with perfect fluid Codazzi tensor since
may be rewritten as in Cotton gravity $R_{kl}-\frac{1}{2}Rg_{kl}=T_{kl}+\mathscr{C}_{kl}-g_{kl}\mathscr{C}_{r}^{r} $,
with 
\begin{equation}
\mathscr{C}_{kl}=2\zeta u_{k}u_{l}+\frac{1}{3}g_{kl}(2\zeta-V)\label{eq: Codazzi mimetic}
\end{equation}
It is a perfect fluid tensor with $\mathscr{A}=2\zeta$ and
$3\mathscr{B}=2\zeta-V$ but in general it is not Codazzi. 
Nevertheless, in view of (\ref{eq: cons mimetic final}), it is always 
$$-\frac{\dot{\mathscr{B}}}{H}=\frac{1}{3H}(\dot{V}-2\dot{\zeta})=2\zeta=\mathscr{A}$$
We report Theorem 2.1 in \cite{Mantica23} restricted to the case of
vanishing acceleration:\\ 
\textit{The perfect fluid tensor $\mathscr{C}_{kl}=\mathscr{A}u_{k}u_{l}+\mathscr{B}g_{kl}$
is Codazzi if and only if: 1) $\nabla_j u_k = H(g_{jk}+u_j u_k)$, 2) $\nabla_{j}H=-\dot{H}u_{j}$, 
3) $\nabla_j \mathscr A = - \dot{\mathscr A}u_j $ and $\nabla_j \mathscr B =-\dot{\mathscr B} u_j$, 4) $H= - \dot{\mathscr B}/\mathscr A $.} \\
This can be rephrased as follows:
\begin{prop}\label{11}
The field equation (\ref{eq: Mimetic equations}) of Mimetic gravity 
is the  field equation of Cotton gravity if and only if the space-time
is GRW, $V=V(t)$ and $\zeta=\zeta(t)$.\\
In particular, in a FRW space-time the field equations (\ref{eq: Mimetic equations})
are the Cotton equations.
\end{prop}
%

%In \cite{Sebastiani 17} it is repeatedly stressed that in
%mimetic gravity, differently from other theories,
%the scalar field is constrained.\\ 

%\subsubsection*{Higher order terms in mimetic gravity in FRW background}

Cotton gravity can include other versions of mimetic gravity.
As noted in \cite{Chamseddine 14} (see also the review \cite{Casalino 21})
in order to have viable cosmological perturbations the action (\ref{eq:Lagrang mimetic})
has to include higher derivative terms. For example
it is possible to add $\frac{1}{2}\gamma(\square\phi)^{2}$ being
$\gamma$ a constant. The new field equations are
(eq. 110 in \cite{Casalino 21}): 
\begin{align}
&R_{kl}-\frac{1}{2}Rg_{kl}=T_{kl}+g_{kl}[V(\phi)+\gamma \nabla_{p}\chi \nabla^{p}\phi] \nonumber\\
&\qquad\qquad +2\zeta\nabla_{k}\phi\nabla_{l}\phi-\gamma[\nabla_{k}\phi\nabla_{l}\chi+\nabla_{k}\chi\nabla_{l}\phi]\label{eq:mimetic high order}
\end{align}
where $\square\phi=\chi$. The background is a FRW space-time.
Since $u_{k}=-\nabla_{k}\phi$ we get $3H=-\square\phi$. Moreover,
recalling that $\nabla_{k}H=-\dot{H}u_{k}$, it is $\nabla_{k}\chi=-3\nabla_{k}H=3\dot{H}u_{k}$
and the previous equation rewrites as 
\begin{align}
R_{kl}-\frac{1}{2}Rg_{kl}=T_{kl}+&2(\zeta+3\gamma\dot{H})u_{k}u_{l} \nonumber\\
&+g_{kl}[V(\phi)+3\gamma\dot{H}]\label{eq:mimetic Hig order pf}
\end{align}
Thus we recognize 
\begin{equation}
\mathscr{C}_{kl}=2(\zeta+3\gamma\dot{H})u_{k}u_{l}+\frac{1}{3}g_{kl}(2\zeta-V+3\gamma\dot{H})\label{eq:Codazzi mim high orde}
\end{equation}
This is again perfect fluid, with $\mathscr{A}=2(\zeta+3\gamma\dot{H})$
and $3\mathscr{B}=2\zeta-V+3\gamma\dot{H}$ being in this context  $V=V(t)$ and $\zeta=\zeta(t)$.
The covariant divergence of (\ref{eq:mimetic Hig order pf}) gives
the conservation law 
\begin{equation}
2 [3H\zeta+9\gamma H\dot H +\dot{\zeta}+3\gamma\ddot{H}]u_{l}+\nabla_{l}V+3\gamma\nabla_{l}\dot{H}=0
\end{equation}
Transvecting this with $u^{l}$ gives 
\begin{equation}
\gamma\ddot{H}=2H(\zeta+3\gamma\dot{H})-\frac{2\dot{\zeta}}{3}+\frac{\dot{V}}{3}\label{eq:cons hig order}
\end{equation}
Thus $3\dot{\mathscr{B}}=2\dot{\zeta}-\dot{V}+3\gamma\ddot{H}$
and using (\ref{eq:cons hig order}) it is $\frac{\dot{\mathscr{B}}}{H}=2(\zeta+3\gamma\dot{H})=-\mathscr{A}$
and (\ref{eq:Codazzi mim high orde}) is a Codazzi tensor.\\ We have proven the following 
\begin{prop}
In a FRW space-time the field equations (\ref{eq:mimetic Hig order pf})
are the Cotton equations.
\end{prop}

%%%%%%%%%%%%%%%%%%%%%%%%CHANGE
%A more general Lagrangian containing a term proportional to $(\nabla_{k}\nabla_{l}\phi)^{2}$
%was investigated in \cite{Casalino 18}. In \cite{Nojiri 14} the authors introduced Mimetic $f(R)$ gravity.
A Lagrangian containing also a term proportional to $(\nabla_k \nabla_l \phi)^2$ was investigated
by Casalino et al. \cite{Casalino 18}. Its viability
was tested in the light of the multi messenger detection of the gravitational
wave event GW170817 and its optical counterpart \cite{Casalino 19}. As a result,
the coefficient multiplying this term was shown to be  $<10^{-15}$; thus the term should be suppressed.

In closing, we recall that Nojiri and Odintsov \cite{Nojiri 14} introduced Mimetic $f(R)$ gravity.
%%%%ENDOFCHANGE

\section{Fixing the dark sector}
From the above discussion it is clear that the dark sector is described by the Codazzi terms and
emerges from geometry. In Cotton gravity the term $\mathscr{B}$ remains unfixed,
so that further restrictions are needed.\\
We make a standard cosmological analysis by supposing that the content of energy in $T_{jk}$ 
is from radiation (r) and matter (m): $\mu = \mu_r+\mu_m$ where
$$\mu_r = \frac{\mu_{r,0}}{(a/a_0)^4}, \quad \mu_m = \frac{\mu_{m,0}}{(a/a_0)^3}$$ 
By setting $\frac{8\pi G}{3H_0^{2}}=1/\mu_c$, $\Omega_{r,0}= \mu_{r,0}/\mu_c$, $\Omega_{m,0}=\mu_{m,0}/\mu_c $,
$\Omega_{k,0}=-\frac{R^{\star}}{6H_0^2 a_0^2}$ and $\Omega_\Lambda=\frac{\Lambda}{3H_0^2}$ we get
\begin{align*}
\frac{H^2}{H_0^2} =\frac{\Omega_{r,0}}{(a/a_0)^4}+\frac{\Omega_{m,0}}{(a/a_0)^3}
+\frac{\Omega_{k,0}}{(a/a_0)^2}+\Omega_{\Lambda}+\frac{\mathscr B}{H_0^2}  %\label{eq:Fried dark}
\end{align*}
In terms of red-shift $1+z=a_0/a$ the equation becomes
\begin{align}
\frac{H^2}{H_0^2}= \Omega_{r,0} (1+z)^4 + \Omega_{m,0}(1+z)^3\nonumber \\
+ \Omega_{k,0}(1+z)^2 +\Omega_{\Lambda}
+\frac{\mathscr{B}(z)}{H_0^{2}}\label{eq:Friedmann z}
\end{align}
If $\mathscr{B}=0$ the standard $\Lambda$CDM model is recovered.\\
It is quite remarkable that we only need to assume the presence of
matter and radiation, while the theory provides the term
that can be interpreted as a dark sector. As Harada argued, the dark sector 
appears as a purely geometric effect due to the presence of the Codazzi tensor.

Let's write the condition $ \dot{\mathscr B} = -H\mathscr A$
as a function of the redshift. 
With $\dot{z}=-(1+z)H$ it is $\dot{\mathscr B}= \frac{d\mathscr B}{dz}\dot{z} =-\frac{d\mathscr{B}}{dz}(1+z)H$. 
The condition becomes
\begin{align}
\mathscr A=\frac{d\mathscr{B}}{dz}(1+z) \label{62}
\end{align}
In this representation $\mathscr A$
does not depend on the Hubble parameter.

Now recall eqs.\eqref{MUD} and \eqref{PD}: $\kappa\mu_{D}= 3\mathscr{B}$ and $\kappa p_D = -3\mathscr B- \dot{\mathscr B}/H$.
Suppose that an EoS $p_{D}=w(z)\mu_{D}$ is valid, where 
the parameter $w$ may be redshift-dependent. \\
In general the dark sector is characterized by $w<-1/3$. The regime $-1<w<-1/3$
is usually called ``quintessence'', while the one with $w<-1$ is
called ``phantom''. The consequences of a phantom
energy in the Universe were pointed out in the seminal paper \cite{Caldwell 03}.

The EoS and \eqref{62} imply the equation
$$3\mathscr B (1+w(z)) = (1+z) \frac{d\mathscr B}{dz} $$
with solution
\begin{align}
\mathscr{B}(z) =\mathscr{B}_0 \exp \left [3\int_0^{z}\frac{1+w(z')}{1+z'}dz'\right  ]
%&\mathscr A(z) =3(1+w(z)) \mathscr B(z)
\label{eq:A,B of z}
\end{align}
Inserting this in (\ref{eq:Friedmann z}) we have 
\begin{align}
\frac{H^2}{H_0^2}= \Omega_{r,0}(1+z)^4 + \Omega_{m,0} (1+z)^3 + \Omega_{k,0}(1+z)^2 \nonumber\\
+\Omega_\Lambda+\frac{\mathscr{B}_0}{H_0^2} \exp \left [3\int_0^z \frac{1+w(z')}{1+z'}dz'\right ]  
\end{align}
This is substantially eq.14 in \cite{Bargiacchi 22}
with the difference that here $\Lambda$ is not dynamical. We also note the balance 
\begin{align}
1=\Omega_{r,0}+\Omega_{m,0}+\Omega_{k,0}+\Omega_{\Lambda}+\Omega_{D,0}\label{eq:balance}
\end{align}
where $\Omega_{D,0} = \mathscr B_0/H_0^2 $ is
the present-time dark energy density. \\
This analysis generalizes 
the considerations in \cite{Harada 23 b,Mantica 23 a}. In particular,
if $w(z)=w$ we get the wCDM model with cosmological constant
reviewed in \cite{Bargiacchi 22}:
\begin{align*}
\mathscr B (z)=\mathscr{B}_0(1+z)^{3(1+w)}, \quad
\mathscr A(z) =3(1+w)\mathscr{B} (z)
\end{align*}
Reversing to cosmic time we get
\begin{align}
&\mathscr B (t) = \frac{\mathscr B_0}{(a(t)/a_0)^{3(1+w)}}\\
&\mathscr{A}(t)=3(1+w)\mathscr B (t)
\end{align}
In the case $w=-5/3$ we recover the phantom term typical of conformal
Killing gravity discovered in \cite{Harada 23 a,Harada 23 b,Mantica 23 a}.\\
The Codazzi tensor becomes 
\begin{align}
\mathscr{C}_{kl}=\frac{\mathscr B_0}{(a/a_0)^{3(1+w)}} \left[3(1+w)u_ku_l+g_{kl}\right]
\end{align}

There are many redshift-dependent models that
parametrize the shape of dark energy: they were used for example in
\cite{Bargiacchi 22} to test deviations from the $\Lambda$CDM
model. More recently they were discussed on the base of JWST results \cite{Maldonado 23}. The same parametrizations can be
used to fix $\mathscr{B}$ and $\mathscr A$ using (\ref{eq:A,B of z}).
We recall some of them here.
\subsection{Chevallier-Polarski-Linder (CPL) model}

It is one of the most used redshift-dependent parametrisation, and
was introduced in \cite{Chevallier01,Linder 03}. It supposes that 
$$ w(z)=w_0+w_a\frac{z}{1+z} $$
where $w_0$ is the present time dark energy EoS parameter and the correction describes
its evolution. It features a good behavior at high $z$ and it is linear at low $z$ (see \cite{Adil 23} and
\cite{Maldonado 23} for details). From (\ref{eq:A,B of z}) we obtain
\begin{align}
&\mathscr B(z)=\mathscr{B}_0(1+z)^{3(1+w_0+w_a)}\exp(-\frac{3w_a z}{1+z})\\
&\mathscr A(z)=3\mathscr{B}(z) (1+w_0+\frac{w_a z}{1+z})
\label{eq:CPL model}
\end{align}
The CPL model has a counterpart in the Codazzi parametrization of Cotton gravity.

%\begin{remark} 
%%%%%%%%%%%%%%%%%%%%%%%%%%%%%%%CHANGE
%In \cite{Adil 23} the authors observe
%that the recent data from JWST reveal a very
%large number of massive galaxies at high redshift. This fact poses
%challenges to the standard $\Lambda$CDM model. Based on the CPL model and testing with the new datasets,
%they propose a scenario in which the Dark Sector consists of a negative cosmological
%constant.
In \cite{Adil 23} the authors observe that
the recent data from JWST reveal a very large number of massive galaxies
at high redshift. This fact poses challenges to the standard $\Lambda$CDM
model. Based on the CPL model and testing with the new datasets, they
propose a scenario in which the Dark Sector consists of a negative
cosmological constant. A similar model was considered in \cite{Visinelli 19}.\\
%\end{remark}

\subsection{Jassal-Bagla-Padmanabhan (JPB) model}
In ref.\cite{Jassal 05} Jassal et al. introduce the following expression
for the EoS parameter, claiming that it solves some issues present in the CPL
model (see \cite{Maldonado 23} and references therein):
$$w(z)=w_0+w_a\frac{z}{(1+z)^2}$$
From (\ref{eq:A,B of z}) we easily obtain
\begin{align}
&\mathscr{B}(z)=\mathscr{B}_0(1+z)^{3(1+w_0)}\exp \left [\frac{3}{2}\frac{w_az^{2}}{(1+z)^{2}}\right ]\\
& \mathscr A(z)=3\mathscr{B}(z)\left [1+w_0+\frac{w_az}{(1+z)^{2}}\right ]
\label{eq:JBP model}
\end{align}
Also this model has a counterpart in the Codazzi parametrization
of Cotton gravity, without explicit introduction of dark energy.

\section{Conclusions}
Cotton gravity offers a simple setting to reproduce the
Friedmann equations of well known extended theories. In all cases the dark sector arising from geometry is described by a Codazzi
tensor with the proper choice of a single function.
%dark sector in the form the arbitrary Codazzi tensor. 
We also showed that the recently proposed conformal
Killing gravity is absorbed in Cotton gravity at least for cosmological
FRW space-times; this is also true for Mimetic gravity. The dark sector may be fixed requesting an EoS: this
can accomodate in a unified description the best known redshift dependent
models.
\medskip

\subsection*{Acknowledgement}
We warmly thank the Referee for suggesting to us the interesting result on Generic Gravity Theory now included in
Section 3, which perfectly corroborates our findings.

\subsection*{Data availability}
Data sharing is not applicable to this article as no datasets were generated or analyzed during the current study.
\end{document}